\documentclass[aps,prl,reprint,amssymb,superscriptaddress,floatfix]
{revtex4-1}
\usepackage{graphicx}
\usepackage{amsmath}
\begin{document}
\title{Non-Fickian current enabling the uphill diffusion of impurities 
diffusing by the mechanism of mobile impurity-defect pairs}
\author{V. I. Tokar}
\affiliation{Universit\'e de Strasbourg, CNRS, IPCMS, UMR 7504,
F-67000 Strasbourg, France}
\date{\today}
\begin{abstract}
Equations governing the diffusion mediated by the mobile pairs have
been derived.  In addition to known non-Fickian (n-F) behavior observable
at small spatiotemporal scale under homogeneous defect distribution the
equations predict another type of n-F diffusion that may take place only
in the inhomogeneous case and can be operative at any scale.  In the
limit of slowly varying inhomogeneity the n-F diffusion current has been
identified and shown to be able to induce the uphill diffusion. It has
been argued that the latter should be observable in about one third of
impurities in FCC metals.
\end{abstract}
\maketitle 
The mechanism of diffusion via mobile impurity-defect (I-d) pairs was
developed in Refs.\ \cite{cowern1990,cowern1991,cowern2007} in order to
explain the unusual exponential diffusion profiles of boron impurities in
silicon. Its peculiarity in comparison with the conventional mechanisms
of defect-mediated diffusion is that the impurity within the pair exists
in a mobile state where it may perform not one but many consecutive
migration steps due to the closeness of the diffusion-assisting defect.
The mobile state, however, has a finite lifetime and as a consequence
the distribution of the migration distances is not Gaussian as in
the conventional Fickian diffusion but exponential in agreement with
experiment \cite{cowern1990}. Similar behavior was later found in B-Ge
system \cite{mirabella} and in the diffusion of indium impurities in
the copper (001) surface layer \cite{In-V-attraction}.

Another unusual phenomenon discovered in B-Si system is the uphill
diffusion, that is, the diffusion along the concentration gradient in
contradiction to the first Fick's law \cite{uphill0}. It was noted
that the phenomenon is correlated with the steepness of the defect
concentration gradients and may be enhanced by the presence of the mobile
pairs, though no explicit mechanisms were suggested \cite{cowern2003}.

Deviations from Fick's laws are of considerable interest from
fundamental and practical standpoints which necessitates their thorough
investigation. The aim of the present Letter is to show that in the
systems where the impurity diffusion is underlain by the mechanism
of mobile pairs an inhomogeneous distribution of defects induces a
non-Fickian (n-F) contribution into the diffusion current which is
enhanced by steep gradients of defect concentration. In contrast to the
n-F exponential behavior restricted to short times, nanometer distances,
and sharp impurity concentration profiles \cite{cowern1990} the action of
the n-F contribution can be observable at arbitrary scales and in smooth
profiles.  A most notable manifestation of the n-F current is the uphill
diffusion that may arise under certain conditions. Because the n-F current
should be operative in all systems with the pair diffusion mechanism
the uphill diffusion should also be observable in all such systems.

Central to the mechanism is the diffusion of an individual I-d pair
which in the continuum approximation can be described by the equation
\cite{In/cu(001),cowern1990,II}
\begin{equation}
	\frac{\partial}{\partial t} G_m({\bf r,r}_0,t)= D_m\nabla^2 
G_m({\bf r,r}_0,t) -\epsilon G_m({\bf r,r}_0,t),
	\label{gastel_eq}
\end{equation}
where $G_m$ is the probability density of finding the mobile pair at time
$t$ at point ${\bf r}$ if the pairing took place at point ${\bf r}_0$ at
time zero; $D_m$ and $\epsilon$ are the pair diffusivity and the decay
rate, respectively.  In comparison with the macroscopic diffusion of
impurities which is limited by the concentration of the point defects
which is usually small, the pair diffusion is very fast because the
mediating defect is permanently present within the pair. However, the
mobile defects are also fast diffusers and so quickly escape the pair.
Because of this the pair lifetime $\tau=\epsilon^{-1}$ is short on the
macroscopic scale. The latter can be characterized by the average time
between successive I-d pairings $t_p=g^{-1}$ where $g$ is the I-d pairing
rate \cite{cowern1990}. The pairing rate is limited by the small defect
concentration $c_d$ because $g=O(c_d)$ \cite{cowern1990,I} and so is
also small.

To derive the equation governing the evolution of the impurity profile
$C({\bf r},t)$ we first note that the macroscopic diffusion proceeds on
the time scale $t=O(t_p)\gg\tau$. At this scale the time step $\Delta
t$ in the governing equation can be chosen in such a way that it was
microscopically large but still small at the macroscopic scale:
\begin{equation}
t=O(g^{-1})\gg\Delta t\gg \tau.
\label{delta_t}
\end{equation}
For example, in experiments of Ref.\ \cite{In/cu(001)} the characteristic
interval between the defect jumps was $10^{-8}$~s. During the pair
lifetime the impurity made on average $\sim10$ steps so $\tau$ was of
$O(10^{-7}$~s). The mean time between the impurity pairings, on the other
hand, was $\sim10$~s, so with the choice of $\Delta t=10^{-3}$~s Eq.\
(\ref{delta_t}) can be easily satisfied.  Because of the four orders of
magnitude difference between $\Delta t$ and $\tau$ in this example, the
creation and the decay of the pair are fully confined within $\Delta t$
with the relative error of $O(10^{-4})$. So on the macroscopic time scale
the redistribution of the impurity density due to the pair diffusion looks
as instantaneous.  This means that within $\Delta t$ the distribution
of the impurity due to one I-d encounter can be approximated by the
integral \cite{In/cu(001)}
\begin{equation}
P({\bf r,r}_0)=\epsilon \int_0^\infty G_m({\bf r,r}_0,t)dt.
\label{Pdef}
\end{equation}
The equation satisfied by $P$ can be obtained by integrating Eq.\
(\ref{gastel_eq}) over $t$ and using the initial condition $G_m({\bf
r,r}_0,0)=\delta({\bf r-r}_0)$ as
\begin{equation}
-\lambda^2\nabla^2P({\bf r,r}_0)+P({\bf r,r}_0)
=\delta({\bf r-r}_0),
\label{P_eq}
\end{equation} 
where $\lambda=\sqrt{D_m/\epsilon}$ is the mean migration distance
\cite{cowern1990,In/cu(001)}. Because the pair carries an impurity, the
impurity conservation forbids the pairs to cross the system boundaries
which leads to the zero Neumann boundary condition for the pair current
\begin{equation}
	{\bf n\cdot\nabla} P({\bf r,r}_0)|_{boundary}=0,
\label{BC}
\end{equation}
where ${\bf n}$ is the vector normal to the boundary.

Thus, on the macroscopic time scale the diffusion of substitutional
impurity is seen as its disappearance at some point due to the pairing
with a defect followed by its instant redistribution in the vicinity of
that point according to the probability density $P$ \cite{In/cu(001)}. By
analogy with the second Fick's law the equation governing the diffusion
under the mechanism of mobile pairs can be written down in the form of
the conservation law
\begin{eqnarray}
\frac{\partial}{\partial t} C({\bf r},t)
&=&-g({\bf r},t)C({\bf r},t)\nonumber\\
&+& \int P({\bf r,r}_0) C({\bf r}_0,t)g({\bf r}_0,t)d{\bf r}_0,
\label{the_eq}
\end{eqnarray}
where the first term on the right hand side describes the loss of
impurities by the concentration profile due to the pairings and the second
term its replenishment by the decayed pairs. Under inhomogeneous defect
distribution the pairing rate $g$ depends on time and space variables. The
impurity conservation is obtained by integrating Eq.\ (\ref{the_eq})
over ${\bf r}$ with the use of Eqs.\ (\ref{P_eq}) and (\ref{BC}).

Eq.\ (\ref{the_eq}) effectively generalizes the physical picture of Ref.\
\cite{cowern1990} on the inhomogeneous case.  To see this let us following
Ref.\ \cite{cowern1990} assume that $g$ is constant and consider an
unbounded host. In this case Eqs.\ (\ref{P_eq}) and (\ref{the_eq}) can
be solved with the use of the Fourier transform. The solution of Eq.\
(\ref{P_eq}) reads
\begin{equation}
	P({\bf k}) = (1+\lambda^2 {\bf k}^2)^{-1},
	\label{Pk}
\end{equation}
where ${\bf k}$ is the Fourier momentum. With its use of $P({\bf k})$ the 
solution of Eq.\ (\ref{the_eq}) is easily found as
\begin{equation}
C({\bf k},t)=\exp\left(-gt\left[1-(1+\lambda^2 {\bf k}^2)^{-1}\right]\right)
C({\bf k},t=0).
	\label{Ckt}
\end{equation}
It can be shown (see Appendix in Ref.\ \cite{I}) that in 1D geometry
with the initial delta-function profile $C({\bf k},t=0)=1$ the inverse
Fourier transform of Eq.\ (\ref{Ckt}) coincides with the solution of Ref.\
\cite{cowern1990}. Thus, the physics described by Eq.\ (\ref{the_eq})
in the homogeneous case is the same so Eqs.\ (\ref{P_eq})--(\ref{the_eq})
can be considered as a straightforward extension of the approach of Ref.\
\cite{cowern1990} on the systems with inhomogeneous defect distributions
and arbitrary 3D geometry.

In the homogeneous case the solution exhibits deviations
from Fickian behavior at short time scale and distances of
$O(\lambda)$ which experimentally were in a few nanometer range
\cite{cowern1990,cowern1991,mirabella}.  Under inhomogeneous defect
distribution Eq.\ (\ref{the_eq}) predicts yet another type of deviation
from Fick's laws which in contrast to the exponential profiles may
manifest itself at arbitrary scales. This can be most easily seen under
stationary conditions when $g({\bf r})$ is time-independent.  In the
conventional defect-mediated diffusion the inhomogeneous distribution
of defects will cause the diffusion constant to be position-dependent.
The diffusion current according to the first Fick's law in this case
will be given by the expression
\begin{equation}
	{\bf J}^F=-D({\bf r})\nabla C({\bf r},t).
	\label{JF}
\end{equation}
Because the diffusion is a relaxation process, at large time the
current should vanish which leads to the stationary solution $C_{st}({\bf
r},t\to\infty)=Const$. Eq.\ (\ref{the_eq}), however, predicts different
stationary distribution
\begin{equation}
	C_{st}({\bf r})=Const/g({\bf r}),
	\label{stationary}
\end{equation}
as follows from the symmetry of the Green function $P$ in Eq.\ (\ref{P_eq}) 
in its arguments under the zero Neumann boundary condition Eq.\ (\ref{BC}).

To clarify the discrepancy between the two cases let us consider the limit
of slowly varying $C$ and $g$ when the integral in Eq.\ (\ref{the_eq})
can be approximated by a local expression. Formally this can be achieved
by expanding Eq.\ (\ref{the_eq}) in powers of $\lambda$ but if the
characteristic scale, say, $\Lambda$ of the spatial variation of both
functions is assumed to be much larger then the migration distance, the
expansion will be, factually, in the powers of $\lambda/\Lambda$. This
is implicitly accounted for in the expansion below in the small values
of the spatial derivatives while $\lambda$ is supposed to have its
physical value.

At the distances of $O(\Lambda)$ from the system's boundaries the
kernel $P$ can be approximated by the inverse Fourier transform of Eq.\
(\ref{Pk}) in the unbounded space
\begin{equation}
P({\bf r-r}_0)=\frac{e^{-|{\bf r-r}_0|/\lambda}}{4\pi\lambda^2|{\bf r-r}_0|}.
	\label{Pr}
\end{equation}
With the use of this expression the integral in Eq.\ (\ref{the_eq})
can be calculated to leading orders in $\lambda$ as
\begin{eqnarray}
	\int d{\bf r}_0 P({\bf r}&-&{\bf r}_0) \phi({\bf r}_0,t)
	=\int d{\boldsymbol \xi}\frac{e^{-|{\boldsymbol \xi}|}}
{4\pi|{\boldsymbol \xi}|} \phi({\bf r}+\lambda{\boldsymbol \xi},t)\nonumber\\	
&=&\phi({\bf r},t)+\lambda^2\nabla^2\phi({\bf r},t)+\dots,
	\label{expansion}
\end{eqnarray}
where $\phi=Cg$, ${\boldsymbol\xi}=({\bf r}_0-{\bf r})/\lambda$, and
the terms linear in $\lambda$ as well as the cross-derivatives of the
second order vanish due to the kernel symmetry.

Thus, Eq.\ (\ref{the_eq}) in the small $\lambda/\Lambda$ approximation 
reads
\begin{eqnarray}
	&&\frac{\partial C({\bf r},t)}{\partial t}\simeq\lambda^2
	\nabla^2[C({\bf r},t)g({\bf r},t)]\nonumber\\
	&&=-\nabla\cdot[-D({\bf r},t)\nabla C({\bf r},t)
	-\lambda^2C({\bf r},t)\nabla g({\bf r},t)],
	\label{diff_eq}
\end{eqnarray}
where 
\begin{equation}
	D({\bf r},t)=\lambda^2g({\bf r},t)
	\label{Ddef}
\end{equation}
\cite{cowern1990}. The first term in the brackets on the second line of
Eq.\ (\ref{diff_eq}) is the Fickian diffusion current Eq.\ (\ref{JF})
while the second term is the n-F contribution. It is interesting to
note that equation similar to Eq.\ (\ref{diff_eq}) was used in Ref.\
\cite{mirabella} in simulations of boron diffusion in amorphous silicon
mediated by the dangling bonds which may hint at a similar underlying
mechanism.

With the use of Eq.\ (\ref{Ddef}) the diffusion current from Eq.\
(\ref{diff_eq}) can be cast in the form
\begin{equation}
	{\bf J}=-D({\bf r},t) C({\bf r},t)[\nabla \ln C({\bf r},t) 
	+ \nabla \ln g({\bf r},t)].
	\label{J}
\end{equation}
As is seen, if the logarithmic derivative of $g$ exceeds that of $C$
and has opposite sign the uphill diffusion follows.  The stationarity
condition ${\bf J}=0$ is satisfied by Eq.\ (\ref{stationary}) also in
this approximation.

To asses the strength of the n-F contribution in physical terms, model 1D
diffusion profiles has been simulated using Eq.\ (\ref{diff_eq}) with and
without the n-F term.  The main difficulty poses the problem of finding
explicit expression for $g$. To find $g$ in realistic simulations one
has to solve the problem of the defect capture by the immobile impurity
given the defect distribution profile $c_d({\bf r},t)$.  This in turn
would require knowledge of details of the defect kinetics and of the
I-d interactions. In homogeneous case, however, the problem simplifies
because $c_d$ is constant and $g$ can be found from the Smoluchowski-type
formula \cite{cowern1990,I} according to which $g$ is proportional to
the defect concentration. This should be roughly valid also in the case
of slowly varying $c_d({\bf r},t)$. Because Eq.\ (\ref{diff_eq}) has
been derived under this assumption, we will use approximation $g\propto
c_d$ in our estimates. The diffusion constant in Eq.\ (\ref{J}) can be
assessed as the product of the equilibrium diffusion constant multiplied
by the supersaturation $c_d({\bf r},t)/c^*_d$ ($c^*_d$ is the equilibrium
value) \cite{cowern1990} while the logarithmic derivative of $g$ in this
approximation will depend only on $c_d$.

\begin{figure}
\begin{center}
\includegraphics[viewport = 30 20 300 360, scale = 0.65]{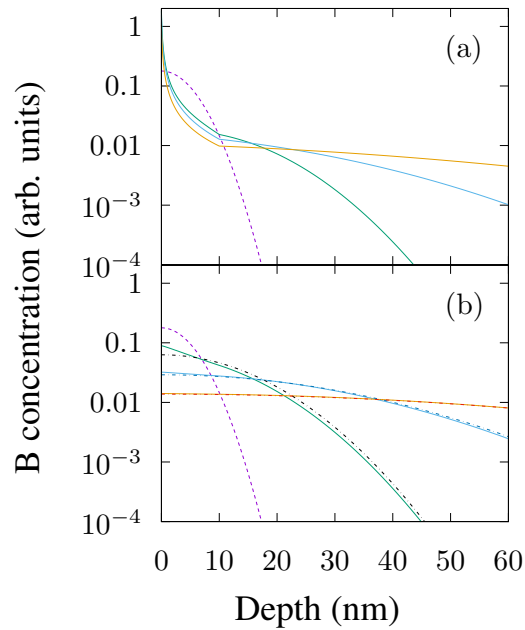}
\end{center}
\caption{\label{FIG1}(Color online) Solid lines: model diffusion profiles
with parameters corresponding to B-Si system simulated with the use
of Eq.\ (\ref{diff_eq}) with (a) and without (b) the n-F term;
the diffusion lasted 60~s at $T=$~850, 900, and 950$^\circ$~C; broader
profiles correspond to higher temperatures. Dashed lines: The initial
profile; dashed-dotted lines: diffusion under homogeneous defect
distribution.}
\end{figure}

For concreteness some characteristic numbers from Refs.\
\cite{cowern1990,supersaturation,cowern2007} were used in the simulations
as the input parameters.  The stationary defect profile was modeled
departing from the fact that in Ref.\ \cite{supersaturation}
strong gradients of the defect concentration were observed at
depth below 10~nm. So $c_d$ was assumed to be constant in Si bulk
with the supersaturation maintained during the experiment, e.\ g.,
by the defect clusters \cite{cowern2007}, at the level of $10^2$
\cite{cowern1990} but starting from the depth 10~nm $c_d$ linearly
dropped to the equilibrium value at the surface \cite{cowern2007}.
The equilibrium diffusion constant was calculated according to Ref.\
\cite{D0Q_B2000}. The results of simulated diffusion shown in Fig.\
\ref{FIG1} are qualitatively similar to the experimental profiles of
Ref.\ \cite{cowern2007}. As is seen, the n-F contribution accounts for
all of the uphill diffusion and the profiles tend to reach the stationary
distribution Eq.\ (\ref{stationary}). The simulated system was assumed
to be unbounded in the positive $x$ direction so at large times the
impurity profile will eventually vanish. In a finite system the steep
decrease of the defect concentration at the other end of the system should
cause the appearance of another peak in the impurity concentration, as
observed experimentally \cite{cowern2007}. If the inhomogeneous defect
distribution persists sufficiently long, at large times the equilibrium
distribution would be reached. Experimentally this may be achieved, e.\
g., with the use of externally imposed constant temperature gradient.

The fact that exponential profiles were seen so far only in a few
systems may be because the values of $\lambda$ in known cases were
restricted to a few nanometers and observations at this scale require
the use of sophisticated experimental techniques. The uphill diffusion,
in contrast, should be realizable at arbitrary length scales which
may considerably increase the number of systems suitable for the study
of the pair diffusion. For this end it would be helpful to establish
some criteria facilitating the search for such systems.  In Refs.\
\cite{cowern1990,cowern1991} it was suggested that the value of $\lambda$
that can be assessed from the impurity diffusion constant (see Eq.\
(\ref{Ddef})) may serve as such a criterion.  Large $\lambda$ means
large number of joint I-d steps, hence, high mobility. If, on the other
hand, $\lambda$ is small and the I-d encounter amounts to only $\sim1$
step the diffusion should be Fickian \cite{cowern1990,mirabella}. Thus,
the pair mechanism should be sought among fast diffusers. This criterion,
however, may be too restrictive in some cases.

For example, in the vacancy-mediated diffusion in the FCC hosts that
microscopically can be described by the five frequency model (5FM)
\cite{LECLAIRE1978} the vacancy can be tightly bound to the impurity but
the latter may still diffuse slowly because of the small value of the
frequency $w_2$ of the I-v (``v'' for vacancy) exchanges that drive the
impurity diffusion.  The strength of the I-v pairing is characterized
in the model by the binding energy $E_b>0$ which can be found from the
detailed balance condition for the associative ($w_3$) and dissociative
($w_4$) jump frequencies as \cite{LECLAIRE1978}
\begin{equation}
	w_3/w_4\simeq\exp(-E_b/k_BT).  \label{Eb}
\end{equation} 
Strong I-v pairing (small decay rate) corresponds to large ratio
$E_b/k_BT$. Thus, for example, in the experiments on Fe-Al system
\cite{voglAlFe} $E_b=0.29$~eV was rather large but $w_2$ was so small that
the number of impurity steps during the pair lifetime was equal to one
(see discussion in Ref.\ \cite{II}).  However, due to the strong binding
and large frequency $w_1$ of the vacancy jumps in the first coordination
shell of the impurity the vacancy makes many random steps around the
impurity before the single diffusive step occurs so the information on
the direction of approach of the defect wipes out.  Correspondingly,
the distribution of the impurity jumps will be symmetric as in the
large-$\lambda$ case (see Eq.\ (\ref{Pr})) which is sufficient for the
derivation of Eq.\ (\ref{the_eq}) and all ensuing equations even in the
case of small $\lambda$.

If, however, the I-v interactions are weak and the diffusion is similar to
the self-diffusion, the one I-v exchange may be only in the direction of
the vacancy that approached the impurity \cite{toroczkai1997}.  So if the
vacancy distribution is inhomogeneous, the diffusive steps will be more
probable toward higher vacancy concentration, that is, in the direction
opposite to that predicted by the n-F term in Eq.\ (\ref{J}). Such
diffusion will also be n-F but in a qualitatively different way.

Thus, the main requirements for the n-F contribution into the diffusive
current of the type of Eq.\ (\ref{J}) is a strong I-v binding, that
is, sufficiently large $E_b$ and sufficiently small temperature. The
binding energies can be estimated with the use of Eq.\ (\ref{Eb})
from the database of Ref.\ \cite{wu_high-throughput_2016} where the
5FM frequencies obtained in {\em ab initio} calculations for $228$ FCC
systems are summarized.  Though the data are not sufficiently reliable
for concrete systems (for example, they predict weak repulsion in the
Fe-Al system instead of the experimentally observed attraction), the
{\em ab initio} calculations are believed to correctly predict general
trends. According to the database $E_b\gtrapprox0.1$~eV in almost third of
the systems so at sufficiently low temperatures the diffusion via mobile
pairs may be sought in more than 70 impurities in FCC metals.

To sum up, in the present Letter the equations governing diffusion of
impurities via the mechanism of mobile pairs have been derived and shown
to exhibit two types of n-F behavior.  The first one is the known behavior
responsible for the exponential diffusion profiles observed at small times
and the nanometer scale distances in boron diffusion in semiconductors. It
was described theoretically in Ref.\ \cite{cowern1990} within the pair
diffusion mechanism under homogeneous defect distributions.  In contrast,
the behavior of the second type is predicted to be observable only in
systems with inhomogeneous distribution of the diffusion mediating defects
and at arbitrary spatiotemporal scales. Its most distinct manifestation
is the uphill diffusion \cite{uphill0} which is predicted to take place
under appropriate conditions in all systems with the diffusion mediated
by the mobile pairs, in particular, in many FCC impurity-host systems.
\begin{acknowledgments}
I would like to express my gratitude to Hugues Dreyss\'e for support.
\end{acknowledgments}
\bibliographystyle{apsrev4-1} 
\end{document}